\documentclass{aa}  
\usepackage{graphicx}
\usepackage{txfonts}
\usepackage[colorlinks=true,
                        linkcolor=blue,
                        citecolor=blue, 
                        filecolor=blue, 
                        urlcolor=blue]{hyperref}
\usepackage[switch]{lineno}
\usepackage{lmodern}
\usepackage{natbib}
\usepackage{graphicx, xspace}
\usepackage{multirow}
\usepackage{amssymb}
\usepackage{subfigure}
\usepackage{lineno}
\usepackage{cancel}

\usepackage{natbib}
\bibpunct{(}{)}{;}{a}{}{,}
\usepackage{amstext}

\begin{document} 

\title{Slow-mode rarefaction and compression fronts in the Hermean magnetosphere: From MESSENGER insights to future BepiColombo observations}

%\subtitle{}
\titlerunning{Slow modes in the Hermean magnetosphere}
\authorrunning{Varela et al.}

   \author{J. Varela\inst{1}
               and  
          F. Pantellini\inst{2}
          }

   \institute{Universidad Carlos III de Madrid, Av. Universidad 30, 28911 Leganes, Spain\\
              \email{\href{mailto:jvrodrig@fis.uc3m.es (telf: 0034645770344)}{jvrodrig@fis.uc3m.es}}
          \and  
             LESIA, Observatoire de Paris, Universit\'e PSL, CNRS, Sorbonne Universit\'e, Universit\'e de Paris, 5 place Jules Janssen, 92195 Meudon, France  \\
             }

\date{version of \today}

% \abstract{}{}{}{}{} 
% 5 {} token are mandatory
 
  \abstract
  {
     \textit{Context:} Standing slow-mode rarefaction and compression front structures may appear in the Mercury magnetosheath under particular solar wind conditions. \\
     \textit{Aims:} The aim of the study is to identify the wind conditions required for the formation of slow-mode structures (SMS) in the Mercury magnetosphere by comparing MESSENGER magnetometer data and magnetohydrodynamics simulations. \\
     \textit{Methods:} We used the magnetohydrodynamics code PLUTO in spherical coordinates to reproduce the interaction of the solar wind with the Mercury magnetosphere. First, the optimal wind conditions for the SMS formation were identified with respect to the orientation of the interplanetary magnetic field (IMF) and dynamic pressure. Next, the magnetic field calculated in the simulations along the MESSENGER trajectory was compared to MESSENGER magnetometer data to identify tracers of the satellite encounter with the SMS. \\     
      \textit{Results:} Optimal wind conditions for the formation of SMS require that the IMF is oriented in the northward or radial directions. The MESSENGER orbit on $8$th September $2011$ takes place during wind conditions that are close to the optimal configuration for SMS formation near the north pole, leading to the possible intersection of the satellite trajectory with the SMS. MESSENGER magnetometer data show a rather strong decrease in the magnetic field module after the satellite crossed nearby the compression front that is observed in the simulation, providing indirect evidence of the SMS.
    }


\maketitle
%\tableofcontents

%\linenumbers
%\modulolinenumbers[5]

\section{Introduction}

Recent space weather studies have shown the important effect of the solar wind (SW) and interplanetary magnetic field (IMF) on the planetary magnetospheres of the Solar System  \citep{Killen,Gonzalez,Vogt}. The SW modifies global magnetospheric structures such as the neighboring magnetosheath, the magnetopause, and the plasma mantle \citep{Cravens}.

Slow-mode structures (SMS) are standing structures that are located between the magnetosheath and the plasma mantle. They are extensively analyzed in the Earth \citep{Zwan,Southwood,Wang} and in Hermean magnetospheres \citep{Pantellini,Varela8}. The SMS expand downstream from the magnetosphere cusp toward the north and south lobes that are integrated in the plasma mantle in the high-latitude nightside magnetosphere, where the SW plasma accelerated by the magnetic reconnections at the magnetopause is partially mirrored \citep{Siscoe,DiBraccio,Jasinski}.

The numerical analysis of SMS in the Earth and Hermean magnetosphere predicts the formation of shocks just upstream of the magnetopause. These shocks are particularly strong in regions with strong magnetic shear near the reconnection points. The SMS are also predicted to form standing structures between the magnetosheath and plasma mantle as a byproduct of the SMS expansion \citep{Taylor}, generating large-scale structures similar to the slow-mode expansion fan defined by \citet{Siscoe} and \citet{Krisko}. Compressional and rarefaction slow-mode fronts, SMCF and SMRF, respectively, were characterized in \citet{Pantellini}. Downstream of a rarefaction front, pressure and density decrease with increasing distance from the front. Downstream of a compressional front, plasma and density increase, while the magnetic field decreases.

Standing structures in the plasma mantle are an important topic because the SW and planetary magnetospheres interact in this magnetospheric region. The SMS might therefore affect the sources and sinks of the inner magnetosphere plasma. The link between SMS with flux transfer events (FTEs) is particularly important in the Hermean magnetosphere because FTEs are an important source of plasma in the mantle \citep{Slavin2,DiBraccio,Jasinski2}.

MESSENGER spacecraft observations revealed several characteristics of the Hermean magnetosphere, for example, a northward shift of the dipolar field by $0.2$ times the planet radius $(R_{M})$, a dipolar moment of $195$nT $R^{3}_{M}$ , and a tilt in the magnetic axis relative to the planetary spin axis smaller than $0.8^{0}$ \citep{Anderson}. Through the magnetometer data that were measured during more than one thousand orbits nearby the northern hemisphere, the Hermean magnetic field can be modeled by an axisymetric multipolar expansion \citep{Richer,Anderson2}. In addition, MESSENGER observations discovered the large variability of the Herman magnetosphere with respect to space weather conditions \citep{Baker5,Anderson3,Johnson}.

The BepiColombo mission will make the orbital insertion around Mercury in $2025$, providing new observational data that will help us to improve the characterization of the Hermean magnetosphere \citep{Benkhoff}. The BepiColombo mission consists of two satellite: the Mercury Planetary Orbiter (MPO), and Mio (the Mercury Magnetospheric Orbiter, MMO) \citep{Milillo}. The instruments on board the MPO and Mio satellites will obtain unique measurements that will complement and extend MESSENGER observations.

The effect of space weather on planetary magnetospheres can be analyzed using different computational frameworks, for example, a single fluid \citep{2015JGRA..120.4763J,Strugarek2,Strugarek}, multifluid \citep{2008JGRA..113.9223K}, and hybrid codes \citep{Muller2012666,Richer,Turc}. The simulations reproduce the bow-shock compression as the SW dynamic and IMF magnetic pressure increase, as well as the effect of the IMF orientation and intensity on the magnetosphere topology \citep{Slavin,2000Icar..143..397K,2009Sci...324..606S,Varela4,Varela7}. Magnetohydrodynamics simulations of the Earth magnetosphere reproduce the effect of space weather conditions on the bow shock \citep{Samsonov,Mejnertsen}, the magnetosheath \citep{Ogino,Wang}, the magnetopause standoff distance \citep{Cairns2,Wang2}, and the magnetotail \citep{Laitinen,Wang3}. Likewise, similar studies were performed for the Hermean magnetic field, showing how the IMF orientation affects the topology of the magnetosphere \citep{1979JGR....84.2076S,2000Icar..143..397K,2009Sci...324..606S} and the plasma flows toward the planet surface \citep{2003Icar..166..229M,2003GeoRL..30.1877K,2004AdSpR..33.2176K,2007AGUFMSM53C1412T,2010Icar..209...11T}.

The aim of the study is to identify the wind conditions that favor the SMS formation in the Hermean magnetosphere. A set of MHD simulations was performed using realistic wind conditions during different MESSENGER orbits to identify the range of SW dynamic pressure, IMF orientation, and intensity linked to the formation of SMS structures. When this parametric range was clarified, a MESSENGER orbit during optimal wind conditions for the formation of SMS nearby the north pole was selected. Next, simulation results were compared with MESSENGER magnetometer data, reproducing the increase in magnetic field intensity and the decrease in plasma density in the plasma mantle as the satellite orbit penetrates the lobes.

The simulations were performed using the single-fluid MHD code PLUTO in spherical 3D coordinates \citep{Mignone}. The analysis is based on the models developed to study the effect of space weather conditions on the global structures of the Hermean magnetosphere \citep{Varela,Varela2,Varela3,Varela4,Varela5}, extreme space weather conditions in the Earth magnetosphere \citep{Varela7}, and the radio emission from exoplanets \citep{Varela6}. The northward displacement of the Hermean magnetic field is represented by a multipolar expansion \citep{Anderson2}. The parameters of the SW such as density, velocity, and temperature were obtained from the numerical model ENLIL + GONG WSA + Cone SWRC \citep{Odstrcil,Parsons}, and the IMF intensity and orientation from the MESSENGER magnetometer data before the satellite entered the Hermean magnetosphere.

This paper is structured as follows. The simulation model, boundary, and initial conditions are described in section \ref{Model}. The dependence of the wind conditions on the SMS formation are analyzed in section \ref{Scan}. The simulation results and MESSENGER magnetometer data are compared during optimal wind conditions for SMS formation in section \ref{Optimal}. Finally, section \ref{Conclusions} summarizes and contextualizes the main conclusions of the study.

\section{Numerical model}
\label{Model}

The simulations were performed using the ideal MHD version of the open-source code PLUTO in spherical coordinates for the nonresistive and inviscid limit \citep{Mignone}. The model solves the time evolution of a single-fluid polytropic plasma in the nonresistive and inviscid limit. The equations solved are
\begin{equation}
\label{Density}
\frac{\partial \rho}{\partial t} + \mathbf{\nabla} \cdot \left( \rho \mathbf{v} \right) = 0
\end{equation}
\begin{equation} 
\label{Momentum}
\frac{\partial \mathbf{m}}{\partial t} + \mathbf{\nabla} \cdot \left[ \mathbf{m}\mathbf{v} - \frac{\mathbf{BB}}{\mu_{0}} + I \left( p + \frac{\mathbf{B}^{2}}{2\mu_{0}} \right)  \right]^{T} = 0 
\end{equation} 
\begin{equation}
\label{B field}
\frac{\partial \mathbf{B}}{\partial t} + \mathbf{\nabla} \times \left(\mathbf{E} \right) = 0
\end{equation} 
\begin{equation}
\label{Energy}
\frac{\partial E_{t}}{\partial t} + \mathbf{\nabla} \cdot \left[ \left( \frac{\rho \mathbf{v}^{2}}{2} + \rho e + p \right) \mathbf{v} + \frac{\mathbf{E} \times \mathbf{B}}{\mu_{0}}  \right] = 0
.\end{equation}
$\rho$ is the mass density, $\mathbf{m} = \rho \mathbf{v}$ is the momentum density, $\mathbf{v}$ is the fluid velocity, $p$ is the gas thermal pressure, $\mathbf{B}$ is the magnetic field, $E_{t} = \rho e + m^2/2 \rho + B^2/2\mu_{0}$ is the total energy density,  $\mathbf{E} = -(\mathbf{v} \times \mathbf{B}) $ is the electric field, and $e$ is the internal energy. The equation of state of an ideal gas was used in the simulations $\rho e = p / (\gamma - 1)$.

The equations were integrated using a Harten, Lax, Van Leer approximate Riemann solver (hll) associated with a diffusive limiter (minmod). The initial magnetic fields were divergenceless. This condition was maintained during the simulation by a mixed hyperbolic and parabolic divergence-cleaning technique \citep{Dedner}.

The grid had $196$ radial points, $48$ in the polar angle $\theta$ and $96$ in the azimuthal angle $\phi$. The grid poles corresponded to the magnetic poles. The simulation domain was confined within two spherical shells, the inner shell at $R = 0.6 R_{M}$ ($R_{M}$ is the radius of Mercury) and the outer shell at $R = 12 R_{M}$. The simulation characteristic length was $R_{M} =2.44 \cdot 10^{6}$ m (the radius of Mercury) and $V = 10^{5}$ m/s was the simulation characteristic velocity (order of magnitude of the SW velocity). No explicit value of the dissipation was included in the model, hence the numerical magnetic diffusivity regulated the typical reconnection in the slow (Sweet–Parker model) regime. A detailed discussion of the numerical magnetic and kinetic diffusivity of the model is provided in \citet{Varela6}.

Special conditions apply for $R < R_{M}$ , where the Alfv\'en velocity was fixed to $\mathrm{v}_{A} = B / (\mu_{0}\rho)^{1/2} =  2.5 \cdot 10^ {4}$ km/s. The plasma density ($\rho_{in}$) in the inner shell was therefore set to
\begin{equation}
\label{eqn:7}
\rho_{in} = \frac{|B|^{2}}{\mu_{0}v_{A}^{2}}
,\end{equation}
and the plasma pressure was defined with respect to the sound speed ($c_{sw}$) of the SW,
\begin{equation}
\label{eqn:8}
p_{in} = \frac{\rho_{in} c_{sw}^{2}}{\gamma}
,\end{equation}
with $\gamma = 5/3$ the polytropic index and $c_{sw} = \sqrt {\gamma p_{sw}/\rho_{sw}} = \sqrt{2\gamma k_{B} T_{sw}/m_{p}}$ the SW sound speed, with $p$ the total electron + proton pressure, $T_{sw}$ the SW temperature and $k_{B}$ the Boltzmann constant. The velocity was smoothly reduced to zero from the planet surface to the inner boundary. In addition, the velocity field was constrained to be parallel to the magnetic field lines.

The outer boundary was divided into the upstream part, in which the stellar wind parameters were fixed, and the downstream part, in which the null derivative condition $\frac{\partial}{\partial r} = 0$ for all fields applied.

The model assumed a fully ionized proton electron plasma. It should be noted that the model does not resolve the plasma depletion layer as a global structure decoupled from the magnetosheath due to a lack of grid resolution. However, the resolution is enough to reproduce global magnetosphere structures such as the magnetosheath and magnetopause, as was demonstrated in previous studies \citep{Varela,Varela2,Varela3}. The magnetic diffusion of the model is higher than that of real plasma, thus, the reconnection between the interplanetary and Hermean magnetic field is continuous (no magnetic pileup on the planet dayside) and stronger (enhanced erosion of the planet magnetic field). Nevertheless, the effect of the reconnection region on the depletion of the magnetosheath and the injection of plasma into the inner magnetosphere was correctly reproduced in a first approximation.

The planetary magnetic field was an axisymmetric potential model with the magnetic potential $\Psi$ expanded in dipolar, quadrupolar, octupolar, and 16-polar terms,
\begin{equation}
\label{eqn:9}
\Psi (r,\theta) = R_{M}\sum^{4}_{l=1} (\frac{R_{M}}{r})^{l+1} g_{l0} P_{l}(cos\theta)
.\end{equation}
The current-free magnetic field is $B_{M} = -\nabla \Psi $. $r$ is the distance to the planet center and $\theta$ is the polar angle. The Legendre polynomials of the magnetic potential are
\begin{equation}
\label{eqn:10}
P_{1}(x) = x
\end{equation}
\begin{equation}
\label{eqn:11}
P_{2}(x) = \frac{1}{2} (3x^2 - 1)
\end{equation}
\begin{equation}
\label{eqn:12}
P_{3}(x) = \frac{1}{2} (5x^3 - 3x)
\end{equation}
\begin{equation}
\label{eqn:13}
P_{4}(x) = \frac{1}{2} (35x^4 - 30x^2 + 3)
.\end{equation}
The numerical coefficients $g_{l0}$ taken from \citet{Anderson2} are summarized in Table \ref{Table 2}.

\begin{table}[h]
\centering
\begin{tabular}{c | c c c c}
coeff & $g_{01}$(nT) & $g_{02}/g_{01}$ & $g_{03}/g_{01}$ & $g_{04}/g_{01}$  \\ \hline
 & $-182$ & $0.4096$ & $0.1265$ & $0.0301$ \\
\end{tabular}
\caption{Multipolar coefficients $g_{l0}$ for the internal field of Mercury.}
\label{Table 1}
\end{table}

It should be noted that the model does not include the effect of the nearly perfectly conducting iron core in the interior of Mercury, which is required to reproduce the consequences of the induction currents on the Hermean surface \citep{Slavin3,Slavin4}. The induction currents lead to an increase in SW dynamic pressure that is required to compress the Hermean magnetopause onto the planet surface \citep{Slavin3,Jia2}. The range of the wind conditions we analyzed is well below the SW dynamic pressure and IMF intensity of an ICME, and therefore, the simulations results provide a reasonable approximation of the global magnetosphere structures in Mercury, even though the effect of the induction currents is not included. 

The numerical magnetic diffusivity regulates the typical reconnection in the slow (Sweet–Parker model) regime because no explicit value of the dissipation was included (a further discussion of the numerical magnetic and kinetic diffusivity of the model is provided in \citet{Varela6}). Consequently, the magnetic diffusion of the model is higher than that of the real plasma, leading to an almost instantaneous reconnection between the interplanetary and Earth magnetic field as well as a rather weak magnetic pileup on the planet dayside. In addition, the reconnection is stronger, leading to an enhanced erosion of the planet magnetic field. The model does not resolve the plasma depletion layer as a decoupled global structure from the magnetosheath because the required resolution is lacking. Nevertheless, the model is able to reproduce the global magnetosphere structures as the magnetosheath and magnetopause, as well as the effect of the reconnection region on the depletion of the magnetosheath and the injection of plasma into the inner magnetosphere in the first approximation. This means that the simulations can reproduce SMS in the tail lobes, although SMS in the dayside are not clearly observed.

The simulations were performed with respect to the Mercury Solar Orbital (MSO) coordinate system: The z-axis is given by the north ecliptic pole, the x-axis is oriented from the center of Mercury to the Sun (antiparallel to the SW flow), and the y-axis is in the duskward direction. The IMF orientations we analyzed are northward ($z>0$), southward ($z<0$), Mercury-Sun ($x>0$), Sun-Mercury ($x<0$), duskward ($y>0$), and dawnward ($y<0$).

The IMF and SW parameters were fixed during the simulation, and the run was completed after steady state was reached after $t = R_{M} / V \approx 4$ min of physical time. This means that the dynamic events caused by changing space weather conditions on timescales on the order of minutes or shorter were not included in the study, for example, substorms \citep{Slavin5,Sun}.

\begin{figure}[h]
\centering
\includegraphics[width=\columnwidth]{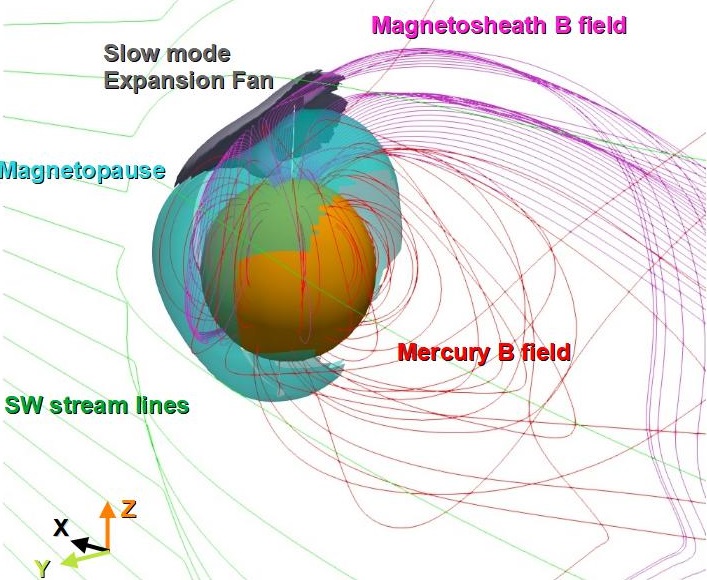}
\caption{3D view of the simulation result corresponding to the $2011/09/08$ MESSENGER orbit. The Hermean magnetic field (red lines), SW stream lines (green lines) and Mercury north pole magnetic field lines are connected with the magnetosheath (pink lines). We also show the magnetopause (cyan isocontour showing $B / B_{IMF} < 1.5$) and the slow-mode expansion fan (gray isocontour showing $p/p_{IMF} > 2.5$).}
\label{fig0}
\end{figure}

Figure \ref{fig0} shows a 3D view of the simulation result corresponding to the $2011/09/08$ MESSENGER orbit. The red lines indicate the magnetic field lines of Mercury in the day- and nightsides. The magnetic field at the dayside is compressed due to the combined effect of the SW dynamic pressure and IMF magnetic pressure, although the magnetic fields on the nightside are stretched, which shapes the magnetotail. The pink lines show the magnetic field lines connecting the Hermean north pole with the magnetosheath. The cyan isocontour indicates the magnetopause location nearby Mercury, and the gray isocontour shows the slow-mode expansion fan. The green lines indicate the SW stream lines that are deflected by the Hermean magnetosphere nearby Mercury.

\section{Optimal wind conditions for the formation of SMS}
\label{Scan}

This section is dedicated to an analysis of SMS in the Hermean magnetosphere for different wind conditions. A set of simulations was performed during MESSENGER orbits in which the IMF was primarily oriented along the axis of the MSO coordinate system. The simulation results provide information about the optimal IMF orientation for the formation of SMS. Table \ref{Table 2} provides the IMF and SW parameters in the simulations.

\begin{table*}[h]
\centering
\begin{tabular}{c | c c c c c c}
IMF & Date & $B_{IMF}$ & $n$ & $T$ & $v$ \\ 
Orientation & Y/M/D & (nT) & (cm$^{-3}$) & ($10^{4}$ K) & (km/s)  \\ \hline
 Mercury-Sun & 2012/01/19 & $(20,0,0)$ & $15$ & $8.5$ & $320$  \\
 Sun-Mercury & 2011/10/17 & $(-18,0,0)$ & $20$ & $9$ & $300$ \\
 Duskward & 2012/03/24 & $(-5,15,5)$ & $30$ & $10$ & $400$ \\
 Dawnward & 2012/03/03 & $(8,-17,0)$ & $70$ & $11.5$ & $380$ \\
 Northward & 2011/09/06 & $(0,-10,41)$ & $90$ & $11$ & $350$ \\
 Southward & 2011/09/29 & $(8,10,-26)$ & $30$ & $6$ & $360$ \\
\end{tabular}
\caption{Wind configuration of MESSENGER orbits in which the IMF was primarily oriented along the axis of the MSO coordinate system. IMF orientation (first column), orbit date (second column), IMF components (third column), SW density (fourth column), SW temperature (fifth column), and SW velocity (sixth column).}
\label{Table 2}
\end{table*}

\begin{figure}[h]
\centering
\includegraphics[width=\columnwidth]{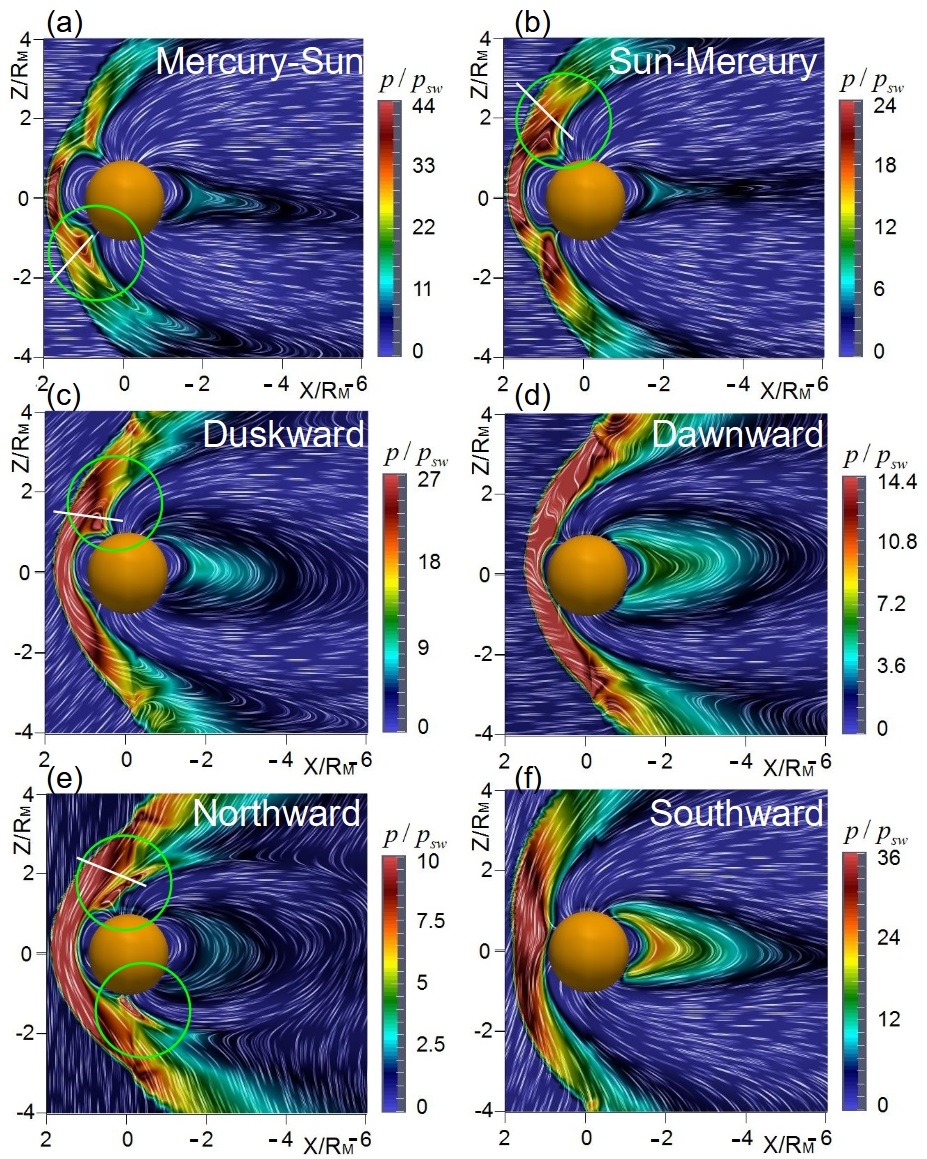}
\caption{Polar cut of the pressure distribution and magnetic field structure (dashed white lines) in simulations with different IMF orientations. (a) Mercury-Sun, (b) Sun-Mercury, (c) duskward, (d) dawnward, (e) northward, and (f) southward IMF orientations. The green circles indicate the location of SMS. The solid white lines show the magnetosphere region analyzed in figure 2.}
\label{fig1}
\end{figure}

Figure \ref{fig1} shows a polar cut of the pressure distribution and the magnetic field structure (dashed white lines). The M-S and northward IMF orientations lead to the formation of SMS near the north pole of Mercury, while the S-M orientation is observed close to the south pole. The southward IMF case shows a strong reconnection between the IMF and the Hermean magnetic field in the equatorial region of the bow shock, which prevents the formation of the SMS because the magnetopause is located too close to the planet surface. The duskward and dawnward IMF orientations do not show clear features of SMS formation, probably due to the east-west tilt induced in the Hermean magnetosphere. Nevertheless, when the duskward and dawnward simulations are compared, the role of the SW dynamic pressure in SMS formation can be analyzed. The SW dynamic pressure is about twice higher in the dawnward simulation than in the duskward case, leading to a smaller standoff distance and a smaller region with closed magnetic field lines on the planet dayside. Consequently, only rather small SMS structures are observed near the north pole in the duskward simulation, but not in the dawnward case. This means that the formation of SMS requires the SW dynamic pressure to stay below some threshold to allow for the magnetopause not to be pushed down to the surface.

\begin{figure}[h]
\centering
\includegraphics[width=\columnwidth]{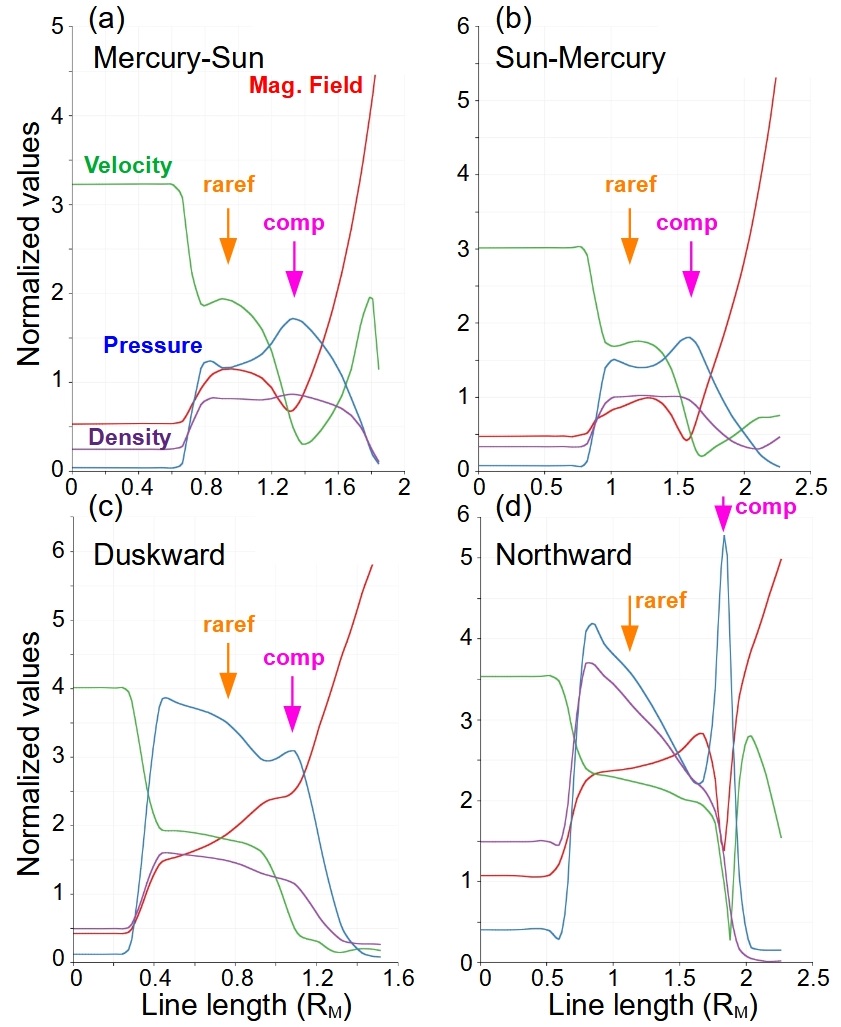}
\caption{Profiles of the magnetic field module, plasma velocity, density, and pressure across SMS in simulations with different IMF orientations. (a) Mercury-Sun, (b) Sun-Mercury, (c) duskward, and (d) northward IMF orientations. The pink arrow indicates the location of the compression front, and the orange arrow shows the rarefaction wave.}
\label{fig2}
\end{figure}

Figure \ref{fig2} shows the module of the magnetic and velocity fields and of the density and pressure along the reference white lines crossing the SMS in figure \ref{fig1}. The profiles are approximately along lines that stand normal to the isobaric surfaces in order to emphasize any anticorrelated variation of the pressure (or the density) and the magnetic field strength. The compression front is linked to a region of maximum field line bending (high pressure), a minimum of the magnetic field module (magnetic reconnection), and a sharp drop in plasma velocity. The rarefaction wave is identified as the region in which the magnetic field module increases while the plasma density and pressure drop. This is observed for the M-S, S-M, and northward IMF orientations, but not for the duskward case, because the local maximum of the pressure is only correlated with a local flattening of the magnetic field. This indicates that the compression front is weak. The case with the northward IMF shows the strongest gradients at the compression front and the widest rarefaction wave. 

In summary, the optimal conditions for the observation of SMS along a MESSENGER trajectory requires a rather low SW dynamic pressure. In addition, IMF orientation should be in the M-S and northward IMF directions.

\section{Signs of SMS in MESSENGER magnetometer data}
\label{Optimal}

This section is dedicated to an analysis of a MESSENGER orbit during wind conditions close to the optimal configuration for the formation of SMS. The orbit during $2011/09/08$ was exposed to a rather intense IMF with dominant components in the S-M and northward directions and an SW dynamic pressure of $5.2$ nPa. The wind parameters are summarized in Table \ref{Table 3}.

\begin{table}[h]
\centering
\begin{tabular}{ c c c c c c c c}
Date & B field& $n$ & $T$ & $v$  \\ 
Y/M/D & (nT) & (cm$^{-3}$) & (K) & (km/s)  \\ \hline
2011/09/08 & $(-30,26,30)$ & $48$ & $155.000$ & $360$ \\
\end{tabular}
\caption{Simulation parameters for the $2011/09/08$ MESSENGER orbit.}
\label{Table 3}
\end{table}

\begin{figure}[h!]
\centering
\includegraphics[width=\columnwidth]{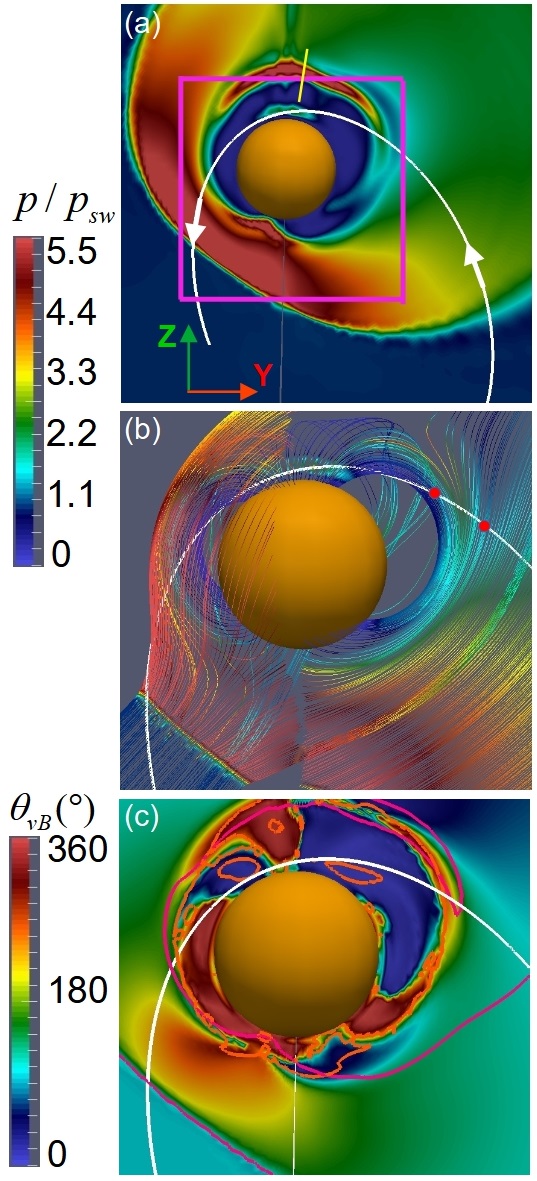}
\caption{Magnetosphere cut in the plane of the satellite trajectory. (a) Plasma pressure distribution in the plane of the satellite trajectory (solid white line). The white arrows indicate the satellite displacement orientation along the orbit. The pink square shows the zoom-in region plotted in the other panels. The yellow line crossing the local maximum of the plasma pressure indicates the magnetosphere region we analyzed in panel 6b. (b) Magnetic field lines connected to the satellite trajectory. The magnetic field line colors indicate the plasma pressure (same scale as in panel a). The red dots show MESSENGER access and exit of the SMS. (c) Angle between the plasma velocity and the magnetic field. The pink isoline indicates the magnetosphere region in which the pressure is $2.5$ higher than the SW pressure. As a reference, the orange contour indicates the loci in which the plasma flow speed is equal to phase speed of the slow mode with antiparallel $k$ vector.}
\label{fig3}
\end{figure}

Figure \ref{fig3} shows SMS in the plane of the satellite trajectory. Panel a indicates the pressure distribution, where possible wide SMS structures are observed nearby the north pole, intersected by the MESSENGER trajectory. Panel b shows the magnetic field lines connected to the MESSENGER trajectory and the fraction of the satellite orbit inside the SMS, represented by the red dots (satellite access and exit of the SMS). The colors of the magnetic field lines indicate the plasma pressure (same scale as in panel a). Panel c shows the angle between the plasma velocity and the magnetic field ($\theta_{vB}$). The magnetic field rotates strongly with respect to the plasma velocity as the satellite trajectory crosses the magnetopause (orange-yellow). The pink isolines indicate the magnetosphere region in which the plasma pressure is $2.5$ times higher than the SW pressure. The pink isoline overlaps with the SMCF, with the antiparallel plasma flow region, and with the MESSENGER orbit, predicting that the satellite may cross the compression front. Standing SMS can only form in regions in which the phase speed of the mode equals the flow speed. For the particular mode propagating antiparallel to the flow (i.e., wave vector $k$ $||$ $v$), the phase speed (in the plasma frame) is
$$ v_{\phi}^{2} = \frac{1}{2} \Big\{ c_{s}^{2} + c_{A}^{2} - \big[ ( c_{s}^{2} - v_{A}^{2})^{2} + 4c_{s}^{2} v_{A}^{2} sin^{2}(\theta_{vB}) \big]^{1/2} \Big\}. $$
Thus, the conditions for this particular mode to be standing is  $v/v_{\phi}(\theta_{vB})=1$. The corresponding contour is shown as a reference in Fig. 4c (orange contour). Slow modes propagate faster (slower) at angles $\theta_{kB}$ smaller (larger) than $\theta_{vB}$ , so that the orange contour loosely indicates regions in which  a slow-mode structure could exist.

\begin{figure}[h]
\centering
\includegraphics[width=\columnwidth]{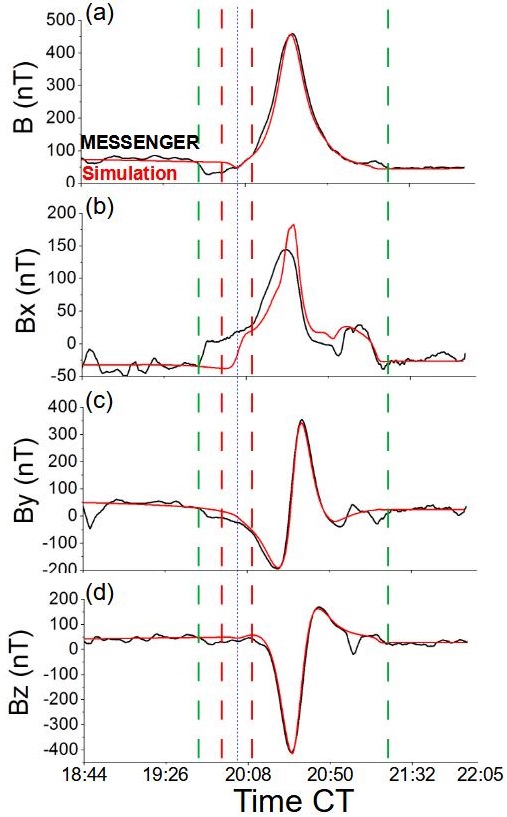}
\caption{MESSENGER magnetometer data vs. simulation magnetic field along the satellite trajectory. (a) Magnetic field module, (b) $B_{x}$, (c) $B_{y}$ , and (d) $B_{z}$ components. The dashed red lines indicate the fraction of the satellite trajectory that is inside the SMS in the simulation. The dotted blue line shows the satellite crossing of the compression front in the simulation. The dashed green lines show the satellite access and exit of the magnetosphere.}
\label{fig4}
\end{figure}

Figure \ref{fig4} compares the magnetometer MESSENGER data (black line) and the magnetic field of the simulation along the satellite trajectory (red line). The simulations provide a reasonable reconstruction of MESSENGER observations. The largest difference is observed in the radial component of the magnetic field (panel b). MESSENGER measured a large decrease in the radial component before the closest approach of the satellite, although the simulation shows an inversion of the radial component, leading to a discrepancy of about $30$ nT. This difference can be explained by a stronger compression of the magnetosphere in the simulations than in the real case, which is probably due to an overestimation of the SW dynamic pressure. The dashed vertical red lines indicate the fraction of the satellite orbit inside the SMS structures, and the dotted blue line shows the location of the compression front based on the simulation results. MESSENGER data show a decrease in magnetic field strength after the satellite enters the magnetosphere (panel a), highlighted by the dashed green line that indicates the satellite bow-shock crossing. The local minimum of the magnetic field strength is displaced farther away from the closest approach of the satellite with respect to the simulation data. The discrepancy is caused by the overestimation of the magnetosphere compression in the simulation. This result might indicate that the fraction of the satellite orbit that is inside the SMS is even larger than in the simulation, showing a steeper decrease in the magnetic field mode and a wider rarefaction wave.

\begin{figure}[h]
\centering
\includegraphics[width=\columnwidth]{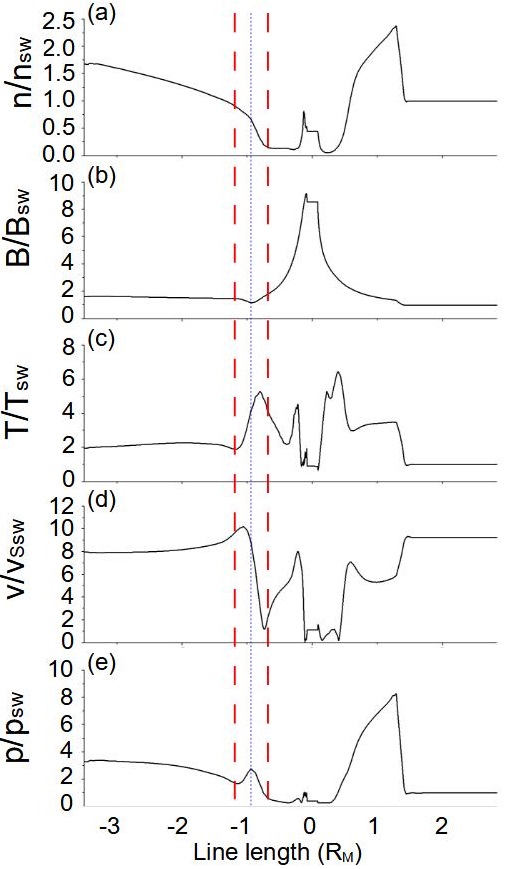}
\caption{Plasma and magnetic field profiles along the satellite trajectory in the simulation. (a) Density, (b) magnetic field module, (c) temperature, (d) velocity module, and (e) pressure. The dashed red lines indicate the fraction of the satellite trajectory that is inside the SMS candidate. The dotted blue line shows the encounter of the satellite with the compression front.}
\label{fig5}
\end{figure}

Figure \ref{fig5} shows the density (panel a), magnetic field module (panel b), temperature (panel c), velocity module (panel d), and pressure (panel e) calculated in the simulation along the satellite trajectory. The dotted blue line indicates the location of the compression front, which is correlated with a sharp decrease in plasma density and velocity module as well as with a local maximum of the plasma temperature and pressure, alongside a local minimum of the magnetic field module. The local maximum of the temperature is displaced with respect to the compression front and is located closer to the closest approach of the satellite, where the local minimum of the plasma velocity is also observed. This indicates that the plasma is heated and slowed down near the compression front. On the other hand, the rarefaction wave is barely observed.

\begin{figure}[h]
\centering
\includegraphics[width=\columnwidth]{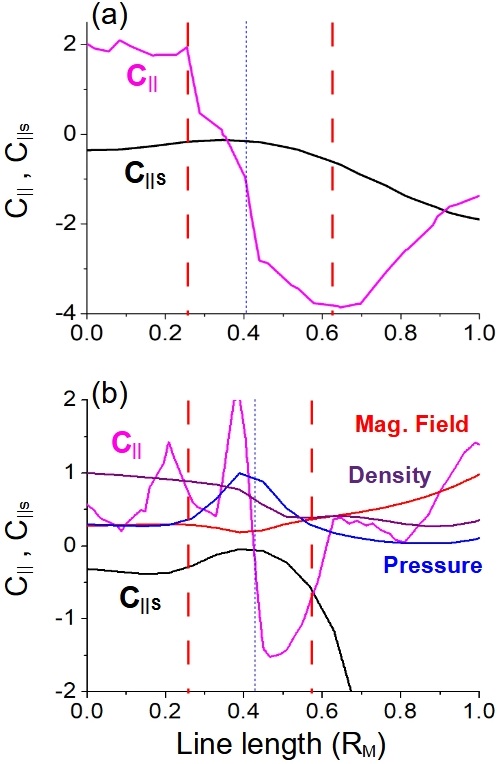}
\caption{Parallel compressibility ($C_{||}$, black line) and parallel compressibility ($C_{||S}$, pink line) calculated using the simulation data. (a) MESSENGER trajectory and (b) straight line indicated in fig. 3a crossing the local maximum of the plasma pressure inside the SMS. The dashed red lines indicate the region with a putative presence of slow modes (the parallel compressibility becomes negative in the immediate vicinity of the magnetopause). The dotted blue line shows the encounter of the satellite with the compression front. Panel b also includes the normalized values of the magnetic field (red), plasma pressure (blue), and density (purple).}
\label{fig6}
\end{figure}

Figure \ref{fig6} shows the parallel compressibility ($C_{||}$, black line) and the parallel compressibility ($C_{||S}$, pink line) calculated using the simulation data along the MESSENGER trajectory (panel a) and a straight line (see fig. 3a) crossing the plasma pressure maximum inside the SMS (panel b). The normalized values to magnetic field, density, and pressure are also included in panel b to indicate the location of the SMS with respect to the parallel compressibility values. The definitions of the parallel compressibility are discussed in Appendix A. As $C_{||}>0$ for the fast mode, the regions with $C_{||} < 0$ are very likely structured by the slow mode. The $C_{||}$ profile sharply decrease after the satellite trajectory crosses the SMS, leading to $C_{||}$ values that are lower than $C_{||S}$ near the SMCF. The same trend is observed for $C_{||}$ and $C_{||S}$ in the straight path along the plasma pressure maximum. The value of $C_{||} > 0$ and is higher than $C_{||S}$ as soon as the straight path exits the SMS. On the other hand, along the MESSENGER orbit, $C_{||}<0$ and takes longer to exceed $C_{||S}$ after the satellite leaves the the SMS. This might indicate that the MESSENGER orbit crosses the tail of the SMS, rather far away from the local maximum of the plasma pressure, where the SMCF shows a sharper transition between the fast and slow mode (see fig. 3a).

In summary, the simulation shows the formation of SMS nearby the Mercury´ north pole on $8$ September 2011. MESSENGER may have crossed an SMS during a fraction of the orbit. MESSENGER magnetometer data indicates a decrease in the magnetic field after the satellite entered the magnetosphere, close to the region in which the simulation identified the SMS.

\section{Discussion and conclusions}
\label{Conclusions}

The SW and the IMF can induce large distortions of the Hermean magnetosphere. These topological variations in the magnetosphere can be analyzed using MHD models to study how the magnetospheric global structures change with respect to the wind conditions.

The presence of SMS in the Hermean magnetosphere is a consequence of the interaction between SW and IMF with the magnetic field of Mercury. Some theoretical analyses were performed to study the properties of SMS in the Hermean magnetosphere \citep{Pantellini,Varela8}, although no evidence of SMS formation was provided. We proposed an indirect method for identifying SMS by combining simulations and MESSENGER magnetometer data.

A set of simulations was performed using realistic wind conditions during MESSENGER orbits with the dominant component of the IMF oriented in different directions. These simulations revealed that radial Mercury-Sun and Northward IMF orientations such as the IMF configurations favor the formation of SMS near the north pole of Mercury. The simulations also showed that if the SW dynamic pressure is too high or the IMF reconnection with the Hermean magnetic field in the planet dayside is too strong (the case of the southward IMF orientation) for a magnetopause  to build up on the dayside, no slow-mode structure is expected to form there because of the critical role of the magnetopause in structuring the plasma flow and the associated standing modes in the magnetosheath. It should be noted that an upper limit of the SW dynamic pressure around $6$ nPa was calculated for the formation of the SMS in the Hermean magnetosphere, even for IMF orientations in the radial Mercury-Sun and northward direction \citep{Varela8}.

After we identified the optimal wind conditions for SMS formation, the MESSENGER magnetometer and ENLIL + GONG WSA + Cone SWRC model databases were screened to select a satellite orbit that could intersect SMS. This is the case of the MESSENGER orbit during $8$  September  $2011$. The MESSENGER magnetometer shows a rather intense IMF with dominant components in the Mercury-Sun and northward directions. In addition, the ENLIL + GONG WSA + Cone SWRC model predicts an SW dynamic pressure of $5.2$ nPa. The simulation indicates the formation of SMS near the north pole that is intersected by the MESSENGER trajectory.

The SMS are identified in the simulations as a magnetosphere region in which the parallel compressibility is negative. In addition, this magnetosphere region shows a local maximum of the plasma pressure, a strong bending of the magnetic field lines, antiparallel plasma flows with respect to the magnetic field, and a local decrease in the magnetic field strength.

The MESSENGER magnetometer data and magnetic fields calculated in the simulation along the satellite trajectory were also compared and agreed reasonably well. The main discrepancy is observed in the radial component of the magnetic field because the simulation predicts a polarity inversion, although the radial magnetic field approaches zero in the magnetometer data. This difference can be explained by an overestimation of the SW dynamic pressure in the simulation, leading to a stronger magnetosphere compression than in the real case. The MESSENGER data show a local minimum of the magnetic field module near the magnetosphere region in which the simulation indicates the intersection of the satellite orbit with the SMS. This result can be understood as indirect evidence of SMS formation in the Hermean magnetosphere.

The BepiColombo mission, particularly the Mio satellite, will provide magnetometer measurements and in situ observations of the plasma properties by the Mercury Plasma Particle Experiment (MPPE) \citep{Baumjohann,Saito,Murakami}. The combination of the two instruments will facilitate the identification and characterization of SMS in the data. The prediction of the optimal wind conditions for SMS formation in this study may help in that endeavor.

\begin{acknowledgements}
This work was supported by the project 2019-T1/AMB-13648 founded by the Comunidad de Madrid and the European Commission's Seventh Framework Programme (FP7/2007-2013) under the grant agreement SHOCK (project number 284515). Data available on request from the authors.
\end{acknowledgements}

\bibliographystyle{aa}
\bibliography{References}

\begin{thebibliography}{66}
\expandafter\ifx\csname natexlab\endcsname\relax\def\natexlab#1{#1}\fi

\bibitem[{{ Anderson} {et~al.}(2011){ Anderson}, {Johnson}, {Korth},
  {Purucker}, {Winslow}, {Slavin}, {Solomon}, {McNutt}, {Raines}, \&
  {Zurbuchen}}]{Anderson}
{ Anderson}, B.~J., {Johnson}, C.~L., {Korth}, H., {et~al.} 2011, Science, 333,
  1859

\bibitem[{{Anderson} {et~al.}(2008){Anderson}, {Acuña}, {Korth}, {Purucker},
  {Johnson}, {Slavin}, {Solomon}, \& {McNutt}}]{Anderson3}
{Anderson}, B.~J., {Acuña}, M.~H., {Korth}, H., {et~al.} 2008, Science, 321,
  82

\bibitem[{{Anderson} {et~al.}(2012){Anderson}, {Johnson}, {Korth}, {Winslow},
  {Borovsky}, {Purucker}, {Slavin}, {Solomon}, {Zuber}, \& {McNutt
  Jr.}}]{Anderson2}
{Anderson}, B.~J., {Johnson}, C.~L., {Korth}, H., {et~al.} 2012, Journal of
  Geophysical Research: Planets, 117, E00L12

\bibitem[{{Baker} {et~al.}(2013){Baker}, {Poh}, {Odstrcil}, {Arge}, {Benna},
  {Johnson}, {Korth}, {Gershman}, {Ho}, {McClintock}, {Cassidy}, {Merkel},
  {Raines}, {Schriver}, {Slavin}, {Solomon}, {Travníček}, {Winslow}, \&
  {Zurbuchen}}]{Baker5}
{Baker}, D.~N., {Poh}, G., {Odstrcil}, D., {et~al.} 2013, Journal of
  Geophysical Research: Space Physics, 118, 45

\bibitem[{{Baumjohann} {et~al.}(2020){Baumjohann}, {Matsuoka}, {Narita},
  {Magnes}, {Heyner}, {Glassmeier}, {Nakamura}, {Fischer}, {Plaschke},
  {Volwerk}, \& {Zhang}}]{Baumjohann}
{Baumjohann}, W., {Matsuoka}, A., {Narita}, Y., {et~al.} 2020, Space Sci. Rev.,
  216, 1

\bibitem[{{Benkhoff} {et~al.}(2021){Benkhoff}, {Murakami}, {Baumjohann},
  {Besse}, {Bunce}, {Casale}, {Cremosese}, {Glassmeier}, {Hayakawa}, {Heyner},
  \& {Hiesinger}}]{Benkhoff}
{Benkhoff}, J., {Murakami}, G., {Baumjohann}, W., {et~al.} 2021, Space Sci.
  Rev., 217, 1

\bibitem[{{Cairns, Iver H.} \& {Lyon, J. G.}(1996)}]{Cairns2}
{Cairns, Iver H.} \& {Lyon, J. G.} 1996, Geophysical Research Letters, 23, 2883

\bibitem[{{Cravens}(1997)}]{Cravens}
{Cravens}, T.~E. 1997, The magnetosphere (Cambridge University Press),
  343–460

\bibitem[{{Dedner} {et~al.}(2002){Dedner}, {Kemm}, {Kröner}, {Munz},
  {Schnitzer}, \& {Wesenberg}}]{Dedner}
{Dedner}, A., {Kemm}, F., {Kröner}, D., {et~al.} 2002, J. Comput. Phys., 175,
  645

\bibitem[{{DiBraccio} {et~al.}(2015){DiBraccio}, {Slavin}, {Raines},
  {Gershman}, {Tracy}, {Boardsen}, {Zurbuchen}, {Anderson}, {Korth}, {McNutt
  Jr.}, \& {Solomon}}]{DiBraccio}
{DiBraccio}, G.~A., {Slavin}, J.~A., {Raines}, J.~M., {et~al.} 2015,
  Geophysical Research Letters, 42, 9666

\bibitem[{{González Hernández, I.} {et~al.}(2014){González Hernández, I.},
  {Komm, R.}, {Pevtsov, A.}, \& {Leibacher, J.}}]{Gonzalez}
{González Hernández, I.}, {Komm, R.}, {Pevtsov, A.}, \& {Leibacher, J.} 2014,
  Solar Origins of Space Weather and Space Climate (Springer-Verlag New York)

\bibitem[{{Jasinski} {et~al.}(2016){Jasinski}, {Slavin}, {Arridge}, {Poh},
  {Jia}, {Sergis}, {Coates}, {Jones}, \& {Waite Jr.}}]{Jasinski2}
{Jasinski}, J.~M., {Slavin}, J.~A., {Arridge}, C.~S., {et~al.} 2016,
  Geophysical Research Letters, 43, 6713

\bibitem[{{Jasinski} {et~al.}(2017){Jasinski}, {Slavin}, {Raines}, \&
  {DiBraccio}}]{Jasinski}
{Jasinski}, J.~M., {Slavin}, J.~A., {Raines}, J.~M., \& {DiBraccio}, G.~A.
  2017, Journal of Geophysical Research: Space Physics, 122, 153

\bibitem[{{Jia} {et~al.}(2015){Jia}, {Slavin}, {Gombosi}, {Daldorff}, {Toth},
  \& {Holst}}]{2015JGRA..120.4763J}
{Jia}, X., {Slavin}, J.~A., {Gombosi}, T.~I., {et~al.} 2015, Journal of
  Geophysical Research (Space Physics), 120, 4763

\bibitem[{{Jia} {et~al.}(2019){Jia}, {Slavin}, {Poh}, {DiBraccio}, {Toth},
  {Chen}, {Raines}, \& {Gombosi}}]{Jia2}
{Jia}, X., {Slavin}, J.~A., {Poh}, G., {et~al.} 2019

\bibitem[{{Johnson} {et~al.}(2012){Johnson}, {Purucker}, {Korth}, {Anderson},
  {Winslow}, {Al Asad}, {Slavin}, {Alexeev}, {Phillips}, {Zuber}, \&
  {Solomon}}]{Johnson}
{Johnson}, C.~L., {Purucker}, M.~E., {Korth}, H., {et~al.} 2012, Journal of
  Geophysical Research: Planets, 117, E00L14

\bibitem[{{Kabin} {et~al.}(2000){Kabin}, {Gombosi}, {DeZeeuw}, \&
  {Powell}}]{2000Icar..143..397K}
{Kabin}, K., {Gombosi}, T.~I., {DeZeeuw}, D.~L., \& {Powell}, K.~G. 2000,
  Icarus, 143, 397

\bibitem[{{Kallio} \& {Janhunen}(2003)}]{2003GeoRL..30.1877K}
{Kallio}, E. \& {Janhunen}, P. 2003, Journal of Geophysical Research, 30, 1877

\bibitem[{{Kallio} \& {Janhunen}(2004)}]{2004AdSpR..33.2176K}
{Kallio}, E. \& {Janhunen}, P. 2004, Advances in Space Research, 33, 2176

\bibitem[{{Kidder} {et~al.}(2008){Kidder}, {Winglee}, \&
  {Harnett}}]{2008JGRA..113.9223K}
{Kidder}, A., {Winglee}, R.~M., \& {Harnett}, E.~M. 2008, Journal of
  Geophysical Research (Space Physics), 113, A09223

\bibitem[{{Killen} {et~al.}(2004){Killen}, {Sarantos}, \& {Reiff}}]{Killen}
{Killen}, R., {Sarantos}, M., \& {Reiff}, P. 2004, Advances in Space Research,
  33, 1899

\bibitem[{{Krisko} \& {Hill}(1991)}]{Krisko}
{Krisko}, P.~H. \& {Hill}, T.~W. 1991, Geophys. Res. Lett., 18, 1947

\bibitem[{{Laitinen, T. V.} {et~al.}(2005){Laitinen, T. V.}, {Pulkkinen, T.
  I.}, {Palmroth, M.}, {Janhunen, P.}, \& {Koskinen, H. E. J.}}]{Laitinen}
{Laitinen, T. V.}, {Pulkkinen, T. I.}, {Palmroth, M.}, {Janhunen, P.}, \&
  {Koskinen, H. E. J.} 2005, Annales Geophysicae, 23, 3753

\bibitem[{{Massetti} {et~al.}(2003){Massetti}, {Orsini}, {Milillo}, {Mura}, {De
  Angelis}, {Lammer}, \& {Wurz}}]{2003Icar..166..229M}
{Massetti}, S., {Orsini}, S., {Milillo}, A., {et~al.} 2003, Icarus, 166, 229

\bibitem[{{Mejnertsen, L.} {et~al.}(2018){Mejnertsen, L.}, {Eastwood, J. P.},
  {Hietala, H.}, {Schwartz, S. J.}, \& {Chittenden, J. P.}}]{Mejnertsen}
{Mejnertsen, L.}, {Eastwood, J. P.}, {Hietala, H.}, {Schwartz, S. J.}, \&
  {Chittenden, J. P.} 2018, Journal of Geophysical Research: Space Physics,
  123, 259

\bibitem[{{Mignone} {et~al.}(2007){Mignone}, {Bodo}, {Massaglia}, {Matsakos},
  {Tesileanu}, {Zanni}, \& {Ferrari}}]{Mignone}
{Mignone}, A., {Bodo}, G., {Massaglia}, S., {et~al.} 2007, ApJS, 170, 228

\bibitem[{{Milillo} {et~al.}(2020){Milillo}, {Fujimoto}, {Murakami}, {Benkhof},
  {Zender}, {Aizawa}, {Dosa}, {Griton}, {Heyner}, {Ho}, \& {Imber}}]{Milillo}
{Milillo}, A., {Fujimoto}, M., {Murakami}, G., {et~al.} 2020, Space Sci. Rev.,
  216, 1

\bibitem[{{M\"{u}ller} {et~al.}(2012){M\"{u}ller}, {Simon}, {Wang},
  {Motschmann}, {Heyner}, {Schüle}, {Ip}, {Kleindienst}, \&
  {Gringle}}]{Muller2012666}
{M\"{u}ller}, J., {Simon}, S., {Wang}, Y.~C., {et~al.} 2012, Icarus, 218, 666

\bibitem[{{Murakami} {et~al.}(2020){Murakami}, {Hayakawa}, {Ogawa}, {Matsuda},
  {Seki}, { Kasaba}, {Saito}, {Yoshikawa}, {Kobayashi}, {Baumjohann}, \&
  {Matsuoka}}]{Murakami}
{Murakami}, G., {Hayakawa}, H., {Ogawa}, H., {et~al.} 2020, Space Sci. Rev.,
  216, 1

\bibitem[{{Odstrcil}(2003)}]{Odstrcil}
{Odstrcil}, D. 2003, Advances in Space Research, 32, 497

\bibitem[{{Ogino, T.} {et~al.}(1992){Ogino, T.}, {Walker, R. J.}, \&
  {Ashour-Abdalla, M.}}]{Ogino}
{Ogino, T.}, {Walker, R. J.}, \& {Ashour-Abdalla, M.} 1992, IEEE Transactions
  on Plasma Science, 20, 817

\bibitem[{{Pantellini} \& {Griton}(2016)}]{Pantellini2}
{Pantellini}, F. \& {Griton}, L. 2016, Astrophys Space Sci, 361, 1

\bibitem[{{Pantellini} {et~al.}(2015){Pantellini}, {Griton}, \&
  {Varela}}]{Pantellini}
{Pantellini}, F., {Griton}, L., \& {Varela}, J. 2015, Planetary and Space
  Science, 112, 1

\bibitem[{{Parsons} {et~al.}(2011){Parsons}, {Biesecker}, {Odstrcil},
  {Millward}, {Hill}, \& {Pizzo}}]{Parsons}
{Parsons}, A., {Biesecker}, D., {Odstrcil}, D., {et~al.} 2011, Space Weather,
  9, S03004

\bibitem[{{Richer} {et~al.}(2012){Richer}, {Modolo}, {Chanteur}, {Hess}, \&
  {Leblanc}}]{Richer}
{Richer}, E., {Modolo}, R., {Chanteur}, G.~M., {Hess}, S., \& {Leblanc}, F.
  2012, Journal of Geophysical Research (Space Physics), 117, A10228

\bibitem[{{Saito} {et~al.}(2021){Saito}, {Delcourt}, {Hirahara}, {Barabash},
  {Andre}, {Takashima}, {Asamura}, {Yokota}, {Wieser}, {Nishino}, \&
  {Oka}}]{Saito}
{Saito}, Y., {Delcourt}, D., {Hirahara}, M., {et~al.} 2021, Space Sci. Rev.,
  217, 1

\bibitem[{{Samsonov, A. A.} {et~al.}(2007){Samsonov, A. A.}, {Sibeck, D. G.},
  \& {Imber, J.}}]{Samsonov}
{Samsonov, A. A.}, {Sibeck, D. G.}, \& {Imber, J.} 2007, Journal of Geophysical
  Research: Space Physics, 112, A12220

\bibitem[{{Siscoe, G.} {et~al.}(2006){Siscoe, G.}, {Crooker, N.U.}, \& {Clauer,
  C.R.}}]{Siscoe}
{Siscoe, G.}, {Crooker, N.U.}, \& {Clauer, C.R.} 2006, Advances in Space
  Research, 38, 173

\bibitem[{{Slavin} {et~al.}(2009){Slavin}, {Acu{\~n}a}, {Anderson}, {Baker},
  {Benna}, {Boardsen}, {Gloeckler}, {Gold}, {Ho}, {Korth}, {Krimigis},
  {McNutt}, {Raines}, {Sarantos}, {Schriver}, {Solomon}, {Tr{\'a}vn{\'{\i}}{\v
  c}ek}, \& {Zurbuchen}}]{2009Sci...324..606S}
{Slavin}, J.~A., {Acu{\~n}a}, M.~H., {Anderson}, B.~J., {et~al.} 2009, Science,
  324, 606

\bibitem[{{Slavin} {et~al.}(2010){Slavin}, {Anderson}, {Baker}, {Benna},
  {Boardsen}, {Gloeckler}, {Gold}, {Ho}, {Korth}, {Krimigis}, {McNutt},
  {Nittler}, {Raines}, {Sarantos}, {Schriver}, {Solomon}, {Starr}, {Travnicek},
  \& {Zurbuchen}}]{Slavin5}
{Slavin}, J.~A., {Anderson}, B.~J., {Baker}, D.~N., {et~al.} 2010, Science,
  329, 665

\bibitem[{{Slavin} {et~al.}(2014){Slavin}, {DiBraccio}, {Gershman}, {Imber},
  {Poh}, {Raines}, {Zurbuchen}, {Jia}, {Baker}, {Glassmeier}, {Livi},
  {Boardsen}, {Cassidy}, {Sarantos}, {Sundberg}, {Masters}, {Johnson},
  {Winslow}, {Anderson}, {Korth}, {McNutt Jr.}, \& {Solomon}}]{Slavin3}
{Slavin}, J.~A., {DiBraccio}, G.~A., {Gershman}, D.~J., {et~al.} 2014, Journal
  of Geophysical Research: Space Physics, 119, 8087

\bibitem[{{Slavin} \& {Holzer}(1979{\natexlab{a}})}]{Slavin}
{Slavin}, J.~A. \& {Holzer}, R.~E. 1979{\natexlab{a}}, J. Geophys. Res., 84,
  2076

\bibitem[{{Slavin} \& {Holzer}(1979{\natexlab{b}})}]{1979JGR....84.2076S}
{Slavin}, J.~A. \& {Holzer}, R.~E. 1979{\natexlab{b}}, Journal of Geophysical
  Research, 84, 2076

\bibitem[{{Slavin} {et~al.}(2012){Slavin}, {Imber}, {Boardsen}, {DiBraccio},
  {Sundberg}, {Saranto}s, {Nieves-Chinchilla}, {Szabo}, {Anderson}, {Korth},
  {Zurbuchen}, {Raines}, {Johnson}, {Winslow}, {Killen}, {McNutt Jr.}, \&
  {Solomon}}]{Slavin2}
{Slavin}, J.~A., {Imber}, S.~M., {Boardsen}, S.~A., {et~al.} 2012, Journal of
  Geophysical Research: Space Physics, 117

\bibitem[{{Slavin} {et~al.}(2019){Slavin}, {Middleton}, {Raines}, {Jia},
  {Zhong}, {Sun}, {Livi}, {Imber}, {Poh}, {Akhavan Tafti}, {Jasinski},
  {DiBraccio}, {Dong}, {Dewey}, \& {Mays}}]{Slavin4}
{Slavin}, J.~A., {Middleton}, H.~R., {Raines}, J.~M., {et~al.} 2019, Journal of
  Geophysical Research: Space Physics, 124, 6613

\bibitem[{{Southwood} \& {Kivelson}(1992)}]{Southwood}
{Southwood}, D.~J. \& {Kivelson}, M.~G. 1992, Journal of Geophysical Research:
  Space Physics, 97, 2873

\bibitem[{{Strugarek} {et~al.}(2014){Strugarek}, {Brun}, {Matt}, \&
  {R{\'e}ville}}]{Strugarek2}
{Strugarek}, A., {Brun}, A.~S., {Matt}, S.~P., \& {R{\'e}ville}, V. 2014, apj,
  795, 86

\bibitem[{{Strugarek} {et~al.}(2015){Strugarek}, {Brun}, {Matt}, \&
  {R{\'e}ville}}]{Strugarek}
{Strugarek}, A., {Brun}, A.~S., {Matt}, S.~P., \& {R{\'e}ville}, V. 2015, apj,
  815, 111

\bibitem[{{Sun} {et~al.}(2022){Sun}, {Dewey}, {Aizawa}, {Huang}, {Slavin},
  {Fu}, {Wei}, \& {Bowers}}]{Sun}
{Sun}, W., {Dewey}, R.~M., {Aizawa}, S., {et~al.} 2022, Science China Earth
  Sciences, 65, 25

\bibitem[{{Taylor} \& {Cargill}(2001)}]{Taylor}
{Taylor}, M. \& {Cargill}, P. 2001, Journal of Plasma Physics, 66, 239

\bibitem[{{Tr{\'a}vn{\'{\i}}{\v c}ek} {et~al.}(2007){Tr{\'a}vn{\'{\i}}{\v
  c}ek}, {Schriver}, \& {Hellinger}}]{2007AGUFMSM53C1412T}
{Tr{\'a}vn{\'{\i}}{\v c}ek}, P.~M., {Schriver}, D., \& {Hellinger}, P. 2007,
  AGU Fall Meeting Abstracts

\bibitem[{{Tr{\'a}vn{\'{\i}}{\v c}ek} {et~al.}(2010){Tr{\'a}vn{\'{\i}}{\v
  c}ek}, {Schriver}, {Hellinger}, {Her{\v c}{\'{\i}}k}, {Anderson}, {Sarantos},
  \& {Slavin}}]{2010Icar..209...11T}
{Tr{\'a}vn{\'{\i}}{\v c}ek}, P.~M., {Schriver}, D., {Hellinger}, P., {et~al.}
  2010, Icarus, 209, 11

\bibitem[{{Turc, L.} {et~al.}(2015){Turc, L.}, {Fontaine, D.}, {Savoini, P.},
  \& {Modolo, R.}}]{Turc}
{Turc, L.}, {Fontaine, D.}, {Savoini, P.}, \& {Modolo, R.} 2015, Journal of
  Geophysical Research: Space Physics, 120, 6133

\bibitem[{{Varela} {et~al.}(2015){Varela}, {Pantellini}, \&
  {Moncuquet}}]{Varela}
{Varela}, J., {Pantellini}, F., \& {Moncuquet}, M. 2015, Planet. Space Sci.,
  119, 264

\bibitem[{{Varela} {et~al.}(2016{\natexlab{a}}){Varela}, {Pantellini}, \&
  {Moncuquet}}]{Varela4}
{Varela}, J., {Pantellini}, F., \& {Moncuquet}, M. 2016{\natexlab{a}}, Planet.
  Space Sci., 129, 74

\bibitem[{{Varela} {et~al.}(2016{\natexlab{b}}){Varela}, {Pantellini}, \&
  {Moncuquet}}]{Varela2}
{Varela}, J., {Pantellini}, F., \& {Moncuquet}, M. 2016{\natexlab{b}}, Planet.
  Space Sci., 120, 78

\bibitem[{{Varela} {et~al.}(2016{\natexlab{c}}){Varela}, {Pantellini}, \&
  {Moncuquet}}]{Varela3}
{Varela}, J., {Pantellini}, F., \& {Moncuquet}, M. 2016{\natexlab{c}}, Planet.
  Space Sci., 122, 46

\bibitem[{{Varela} {et~al.}(2016{\natexlab{d}}){Varela}, {Pantellini}, \&
  {Moncuquet}}]{Varela8}
{Varela}, J., {Pantellini}, F., \& {Moncuquet}, M. 2016{\natexlab{d}},
  Planetary and Space Science, 125, 80

\bibitem[{{Varela} {et~al.}(2016{\natexlab{e}}){Varela}, {Reville}, {Brun.},
  {Pantellini}, \& {Zarka}}]{Varela5}
{Varela}, J., {Reville}, V., {Brun.}, A.~S., {Pantellini}, F., \& {Zarka}, P.
  2016{\natexlab{e}}, A\&A, 595, A69

\bibitem[{{Varela, J.} {et~al.}(2018){Varela, J.}, {Reville, V.}, {Brun, A.
  S.}, {Zarka, P.}, \& {Pantellini, F.}}]{Varela6}
{Varela, J.}, {Reville, V.}, {Brun, A. S.}, {Zarka, P.}, \& {Pantellini, F.}
  2018, A\& A, 616, A182

\bibitem[{{Varela, J.} {et~al.}(2022){Varela, J.}, {Reville, V.}, {Brun, A.
  S.}, {Zarka, P.}, \& {Pantellini, F.}}]{Varela7}
{Varela, J.}, {Reville, V.}, {Brun, A. S.}, {Zarka, P.}, \& {Pantellini, F.}
  2022, A\& A, 659, A10

\bibitem[{{Vogt} {et~al.}(2019){Vogt}, {Gyalay}, {Kronberg}, {Bunce}, {Kurth},
  {Zieger}, \& {Tao}}]{Vogt}
{Vogt}, M.~F., {Gyalay}, S., {Kronberg}, E.~A., {et~al.} 2019, Journal of
  Geophysical Research: Space Physics, 124, 10170

\bibitem[{{Wang, J. Y.} {et~al.}(2014){Wang, J. Y.}, {Wang, C.}, {Huang, Z.
  H.}, \& {Sun, T. R.}}]{Wang3}
{Wang, J. Y.}, {Wang, C.}, {Huang, Z. H.}, \& {Sun, T. R.} 2014, Journal of
  Geophysical Research: Space Physics, 119, 1887

\bibitem[{{Wang, M.} {et~al.}(2012){Wang, M.}, {Lu, J.}, {Liu, Z.}, \& {Pei,
  S.}}]{Wang2}
{Wang, M.}, {Lu, J.}, {Liu, Z.}, \& {Pei, S.} 2012, Chin. Sci. Bull., 57, 1438

\bibitem[{{Wang, Y. L.} {et~al.}(2004){Wang, Y. L.}, {Raeder, J.}, \& {Russell,
  C. T.}}]{Wang}
{Wang, Y. L.}, {Raeder, J.}, \& {Russell, C. T.} 2004, Annales Geophysicae, 22,
  1001

\bibitem[{{Zwan} \& {Wolf}(1976)}]{Zwan}
{Zwan}, B.~J. \& {Wolf}, R.~A. 1976, Journal of Geophysical Research
  (1896-1977), 81, 1636

\end{thebibliography}

\begin{appendix}

\section{Definition of the parallel compressibility}
 
The parallel compressibility of a plane mode with wave vector $\vec{k}$ is
$$ C_{||} = \frac{\delta n}{n} \frac{B}{\delta B_{||,}}$$
with $\delta n$ and $\delta B_{||}$ the variation in density and magnetic field along the direction parallel to the wave vector. For an arbitrary wave vector, the parallel compressibility can be expressed as
$$ C_{||} = \frac{\vec{k} \cdot \vec{\nabla} n}{n} \frac{B}{\vec{k} \cdot \vec{\nabla} B_{||.}} $$
The parallel compressibility of the slow and fast modes is given by
$$ C_{||}(\theta) = \frac{v_{A}^{2}}{v_{\phi}^{2}(\theta)-c_{s,}^{2}} $$
with $\theta$ the angle between the wave vector and the magnetic field and $v_{\phi}$ the phase speed of the corresponding mode. The compressibility of the fast mode is positive (faster than the sound speed) and the compressibility is negative (slower than the sound speed). For a more detailed description, we refer to \citet{Pantellini2}.
 
\end{appendix}

%\newpage
%\bibliographystyle{aa}
%\bibliography{/Publications/Mercury/Radio emission/A&A/aa}

\end{document}